\documentstyle[12pt]{article}
\RequirePackage{graphics}
\renewcommand{\thesection}{{\small {\bf \Roman{section}.}}}

\begin{document}
\medskip
\begin{center}
{\Large \bf Rummaging through Earth's attic for remains of ancient life}
\bigskip\medskip

{\rm \large John C. Armstrong }

{\it Center for Astrobiology and Early Evolution and the Department of Astronomy,
University of Washington, Box 351580 Seattle, WA 98195 \\
 {\footnotesize Email: jca@astro.washington.edu}}

\vspace*{1cm}

{\rm \large Llyd E. Wells}

{\it Center for Astrobiology and Early Evolution and the School of Oceanography,
University of Washington, Seattle, WA 98195}
\vspace*{1cm}

{\rm \large Guillermo Gonzalez}

{\it Department of Physics and Astronomy, Iowa State University, Ames, IA 50011-3160}
\end{center}
\vspace*{2.5cm}

\begin{tabular}{lr}
Text Pages: & 51 \\
Tables: & 6 \\
Figures: & 4 \\
\end{tabular}

\newpage

\noindent\hspace*{5mm} {\Large Proposed Running Head:} {\large Rummaging through Earth's attic} \\
\bigskip\bigskip

\noindent\hspace*{5mm} {\Large Editorial correspondence to:} \\
\smallskip

\noindent\hspace*{5mm} John C. Armstrong \\
\noindent\hspace*{5mm} Department of Astronomy \\
\noindent\hspace*{5mm} University of Washington \\
\noindent\hspace*{5mm} Box 351580 \\
\noindent\hspace*{5mm} Seattle, WA 98195\\
\noindent\hspace*{5mm} USA \\
\medskip

\noindent\hspace*{5mm} Phone: (206) 543.9039 \\
\noindent\hspace*{5mm} Email: jca@astro.washington.edu \\
\newpage

\bigskip

\noindent{\large \bf ABSTRACT}

We explore the likelihood that early remains of Earth, Mars, and Venus have been preserved on the Moon in high enough concentrations to motivate a search mission. During the Late Heavy Bombardment, the inner planets experienced frequent large impacts. Material ejected by these impacts 
near the escape velocity would have had the potential to land and be 
preserved on the surface of the Moon. Such ejecta could yield information 
on the geochemical and biological state of early Earth, Mars and Venus. In 
order to determine whether the Moon has preserved enough ejecta to 
justify a search mission, we calculate the amount of Terran material incident on the Moon over
its history by considering the distribution of ejecta launched from the Earth by large impacts.  In addition, we make analogous estimates for Mars and Venus. We find, for a well mixed regolith, that the median surface abundance of Terran material is roughly 7 ppm, corresponding to a mass of approximately 20,000 kg of Terran material over a 10 x 10 square km area.  Over the same area, the amount of material transferred from Venus is 1-30 kg, and material from Mars as much as 180 kg.
Given that the amount of Terran material is substantial, we estimate the fraction of this material surviving impact with intact geochemical and biological tracers.  
\bigskip

\newpage

\section{{\small {\bf INTRODUCTION}}}
	The frequency of both lunar and Martian meteorites on the Earth indicates that the transfer of planetary material is common in the solar system. Vigorous hydrologic or tectonic cycles, past or present, prevent most nearby planetary bodies from serving as long-term repositories of this material. The Moon is an important exception, however. Strategically located within the inner solar system, the Moon has theoretically collected material from all of the terrestrial planets since its formation. Lacking an atmosphere and widespread, long-lasting volcanism, the Moon has potentially preserved meteorites from Mercury through the asteroid belt. While the lack of an atmosphere prevents a soft landing on the lunar surface, its low gravity means particles with small velocities with respect to the Moon will experience relatively low impact velocities. Moreover, unlike on other terrestrial planets, Martian, Venusian and Terran meteorites, blasted off their respective planets 3.9 Ga during the Late Heavy Bombardment, should still exist on the surface of the Moon. Such meteorites are likely to contain uniquely preserved remains of these planets that are not available elsewhere in the Solar System. In particular, Terran meteorites on the Moon may provide a substantive geological record for ancient Earth, corresponding to or predating the period for which the earliest evidence for life exists.

	In considering the argument in favor of searching the Moon for Terran meteorites, the SNC meteorites represent obvious analogs. A considerable amount of Martian geoscience is built upon little more than a dozen samples. From them have been inferred important constituents of the atmosphere, mantle and core, the extent of interaction between the Martian hydrosphere and lithosphere, constraints on Martian water abundance, and the nature of a Martian carbon sink, while the lightly shocked condition of the meteorites has stimulated developments in the understanding of impact physics (McSween 1994). Terran meteorites have the potential to provide similar information, extending and broadening Earth's geologic record for a time period that has otherwise left little or no physical evidence. The rocks' elemental composition and mineralogy (in particular, hydration) could be used to constrain characteristics of the early crust and mantle, the global oxidation state, the extent of planetary differentiation, and the availability of water. Volatile inclusions sampling noble gases, carbon dioxide and molecular nitrogen could clarify atmospheric origin and evolution and, along with the meteorite mineralogy, could provide substantive constraints on early atmospheric concentrations (Bogard and Johnson 1983). Direct measurement of the timing, extent and planetary effects of the heavy bombardment by careful dating of Terran meteorites is also possible and would perhaps be the most robust and significant scientific reward of this project. 

	In addition to the scientific benefits listed above, which alone justify searching the Moon for Terran meteorites, a fraction of Earth-derived material on the Moon could contain geochemical and biological information, in the form of isotopic signatures, organic carbon, molecular fossils, biominerals or even, theoretically, microbial fossils. Again, analogy to SNC meteorites is instructive. Questionable interpretations of structures within the Martian meteorite ALH84001 as microbial fossils (McKay {\em et al.} 1996) and, more recently, evidence supporting deposition there of magnetite by biomineralization (Thomas-Keprta {\em et al.} 2001; Friedmann {\em et al.} 2001) have been used to argue that this meteorite contains vestiges of ancient Martian life. While such interpretations remain highly controversial, they support the general principle that Terran meteorites should be examined for potentially novel evidence concerning early Earth life. Such evidence could substantiate or extend a contested fossil record that begins 3.5 Ga (Awramik 1982; Schopf and Packer 1987; Buick 1990; Brasier {\em et al.} 2002) and geochemical evidence from even earlier periods, between 3.7 and 3.85 Ga (Schidlowski 1988; Mojzsis {\em et al.} 1996; Rosing 1999; but see Moorbath 2001 and Fedo and Whitehouse 2002). 

In addition, the Moon may preserve material not only from Earth, but also from Venus and the asteroid belt. The only attainable record of Venus' early surface geology, otherwise catastrophically erased 700 million years ago (McKinnon {\em et al.} 1997), is probably on the Moon. Similarly, a record of the type, characteristics and origins of the heavy bombardment impactors themselves may be available on the Moon. Such a record would clarify not only the geological history of Earth, but also its chemical and biological history -- especially since these impactors were potentially major sources of biotic precursors on early Earth (Pierazzo and Chyba 1999).

	Mars is presently the focus of attention with regard to the search for early signs of life outside Earth. Ironically, the Moon may be the better place to search for the remains of both early Martian and early Terran life. Most significantly, the Moon lacks the water capable of carrying contaminants into the interior of rocks through cracks. While gardening from micrometeorites is less severe on the surface of Mars, the deeply buried regolith on the Moon provides some protection.   Finally, the Moon is also a perfect testbed for targeted sample return.  

	For these reasons, we determined the likelihood that early remains of Earth, Mars, and Venus have been preserved on the Moon 
in high enough concentrations to motivate a search mission.  
While others (Gladman 1997; Halliday {\em et al.} 1989) provide
estimates for the transfer efficiency and total number of Martian
meteorites impacting the Earth, there are no estimates for the
abundance of Terran, Martian, and Venusian meteorites on the Moon.
In Section 2, we consider the transfer of impact ejecta from Earth
to the Moon's surface immediately following an impact event.  By
considering both the slow moving, Earth-bound ejecta and the high
velocity ejecta that achieves orbit around the Sun, we estimate the
total transfer efficiency of the material.  Through a separate
numerical simulation, we estimate a rough transfer efficiency for 
Venusian material.  Additionally, we compute the range of expected impact velocities on the surface of the Moon.
In Section 3 the results of Section 2 will be combined with 
mass flux estimates to 
calculate concentrations of Terran and Venusian meteorites on the
Moon.  In addition, the contribution from Martian meteorites is estimated from
the literature.  
Section 4 includes a discussion of the survivability of the material 
delivered to the Moon via impacts over a range of 
velocities.  

\section{{\small {\bf EJECTA TRANSFER FROM EARTH TO THE MOON}}}

Large impacts on the Earth generate ejecta with 
launch velocities near to or exceeding the escape velocity.  This material
interacts with the Moon in a number
of ways.  Ejecta with speeds near the escape
velocity eventually falls back to the Earth, gets swept up by the Moon, 
or achieves a stable orbit around the Earth.  Ejecta with velocities
exceeding the escape velocity either hits the Moon while escaping the
Earth-Moon system, or achieves a stable solar orbit.  We explore the
velocity distribution of the ejecta from low speeds that just allow 
particles to reach the lunar orbit, up to speeds exceeding the
escape velocity.  We break the velocity distribution into three
regimes:

\begin{enumerate}
\item{Ejecta with launch speeds less than the escape velocity, but
still large enough to reach the lunar
orbit.  The minimum velocity is
determined by the Earth-Moon distance, as discussed below.  This ejecta
has a sufficiently low relative velocity with respect to the Moon that 
gravitational focusing is important.  We refer to this as ``direct
transfer,'' and use geometrical and analytical methods to derive the
amount of material that reaches the Moon via this process.}
\item{Ejecta that leaves the Earth-Moon system without interacting
with the Moon and achieves a solar orbit, with the potential of
interacting with the Earth and Moon at a later time.  This includes all ejecta
with velocities greater than the escape velocity.  We refer to
this regime as ``orbital transfer,'' and use
numerical simulations similar to Gladman 1997 to compute the
transfer efficiencies.}
\item{Ejecta that leaves the Earth with speeds greater than the escape
velocity and just happens to hit the Moon as it leaves the system.  
These high velocity ``lucky shots'' are estimated along with the numerical calculations.}
\end{enumerate}

\subsection{{\small {\bf Analytical treatment of direct transfer}}}

The analysis for direct transfer treats the ejecta as a shell of material emanating
from the surface of the Earth.  This shell is defined by its leading and trailing edges, and the total volume increases with time as the shell rises ballistically from the Earth's surface.  As this volume increases, the mean density of material within the shell is reduced.  When the leading edge reaches the lunar orbit, the shell interacts with the Moon until the trailing edge also reaches the lunar orbit.  Ideally, this entire process then repeats in reverse, ignoring the perturbing influence of the Moon, as the shell free-falls back to the surface.  

It is important to realize the role of the lunar period in this transfer
method.  The transfer of material will be most efficient when the Moon is 
closest to the Earth, at 3.8 - 3.9 Ga.  The density of material in the
shell is determined by the lunar orbit, which was much smaller in
earlier epochs.  In addition, the amount of mass incident
on the Earth during the Late Heavy Bombardment (and thus the mass of ejecta 
thrown up by these impacts) was much higher.  These two effects have the
potential to transfer large amounts of Terran ejecta, as described in
Section 3.

The Moon's present mean distance from the Earth, in units of Earth
radii, $R_{\oplus}$, is $\sim 60 R_{\oplus}$, and it is receding at $3.82 \ cm \ yr^{-1}$ (Dickey {\em et al.} 1994).  
Paleontology gives us constraints on this rate as far back as 2.5 Ga 
(Walker and Zahnle 1986).  However, the rapid lunar recession in the first 
$\sim 700$ Myrs prevents an accurate extrapolation to the late heavy 
bombardment epoch from present and paleontological data.  
A recent study (Zharkov 2000) attempts to estimate an early reference point
for lunar recession.  By assuming that the Moon's present shape 
was ''frozen-in'' during the Late Heavy Bombardment epoch, Zharkov
uses the current data from the lunar gravity field and a model of its
tidal interaction with the Earth to estimate this
reference point.  Zharkov estimates that the Moon's mean distance was $21.6 
R_{\oplus}$ about 3.9 Ga.  We will adopt this value in our calculations, and as the starting point for discussions of the time evolution
of the Moon's orbit subsequent to 3.9 Ga.

The leading edge of the shell is fixed by the escape velocity of the Earth (we consider material with velocities greater than the escape speed in Section 2.2).  The trailing edge, however, is determined by the Earth-Moon distance.  Thus, the maximum velocity is given by $v_{max} = 11.2 \ km \ s^{-1}$ and the minimum velocity, $v_{min}$, is determined by considering conservation of energy in the Earth's potential: 

\begin{equation}
v_{min} = v_{esc} \left( 1 - \frac{1}{\eta} \right)^{\frac{1}{2}},
\end{equation}

\noindent where $v_{esc}$ is the Earth's escape velocity and $\eta = \frac{R_{Moon}}{R_\oplus}$ is the distance between the Earth and the Moon divided by the Earth's radius. 
For the case of $\eta$ = 21.6 at 3.9 Ga, $v_{min} = 10.94 \ km \ s^{-1}$.

The position of the shell is determined by solving the system of equations governed by the Earth's gravitational potential,

\begin{equation}
 \frac{\partial v}{\partial t} = -\frac{GM_{\oplus}}{r^{2}};\ \ \ \
\frac{\partial r}{\partial t} = v,
\end{equation}

\noindent where $G$ is the gravitational constant, $M_{\oplus}$ is the Earth's mass, and $r$ and $v$ are the instantaneous position and velocity of the shell's edge.  This set of equations is solved separately for the leading and trailing edges of the shell.
From this, we can deduce the volume of the hemispherical shell,

\begin{equation} 
V_{shell}  = \frac{2}{3}\pi\left( r_{inner}^{3} - r_{outer}^3\right),
\end{equation}

\noindent where $r_{inner}$ is the distance from the Earth to the trailing edge, and $r_{outer}$ is the distance to the leading edge.  The density of material within the shell, $\rho_{shell}$, is simply the mass ejected with speeds between $v_{min}$ and $v_{max}$ divided by this volume.  

Chyba {\em et al.} 1994 calculate the amount of material leaving the surface with velocity greater than a given velocity, $v$, as

\begin{equation}
M_{e} \left(> v \right) = 0.11 \left( \rho/\rho_t \right)^{0.2}
	\left( v_{i}/v \right)^{1.2} m,
\end{equation}

\noindent where $\rho$ and $\rho_t$ are the density of the impactor and target,
and $v_{i}$ is the impact velocity.  In the Earth's case the 
incident material is
primarily from the asteroid belt and $v \sim 14 \ km \ s^{-1}$ (Chyba {\em et al.} 1994; Bottke {\em et al.} 1994).  Assuming $\rho = \rho_t = 2860 \ kg \ m^{-3}$ for basalt, we derive that $M_e \left( > v_{min}
\right) - M_e \left( > v_{max} \right) = 0.004 \ m$ with $\eta = 21.6$ at 3.9 Ga.  This amount will be reduced as the Moon recedes from the Earth over time.

The Moon will intersect the leading edge at a 45 degree angle, with a relative velocity, $v_{rel}$, determined by the quadrature sum of the leading edge velocity and the lunar orbital velocity.  The incremental path length, $dP$, traversed by the Moon in a time $dT$ is

\begin{equation}
dP = v_{rel}dT,
\end{equation}

\noindent while the incremental mass swept up by the Moon is

\begin{equation}
dM = \rho_{shell} \sigma dP; \ \ \ \ \sigma = A_{m}\left( 1 + \left(\frac{v_{esc,m}}{v}\right)^{2}\right),
\end{equation}

\noindent  where $A_{m}$ is the cross sectional area of the Moon, and $v_{esc,m}$ is the lunar escape velocity.  The gravitationally enhanced cross section, $\sigma$, is then integrated over the normalized velocity distribution, and the incremental mass, $dM$, is integrated from the time the leading edge reaches the Moon to the time the trailing edge reaches it.  We multiply this total mass by a factor of 2 to take into account the reverse process of the shell falling back to the Earth.  Table 1 shows the results of the model for nine positions of the Earth and Moon from 3.9 Ga to the present, corresponding to an Earth-Moon distance from 21.6 to 60.3 $R_{\oplus}$.  In particular, we report $F_{direct}$, the fraction of mass reaching the Moon, as a function of the Earth-Moon distance in terms of the impactor mass.

{\bf [ Table 1 ] }

\subsection{{\small {\bf Numerical simulations of orbital transfer}}}

A second method to transfer material from the Earth to the Moon is orbital transfer, which is much less sensitive to the Earth-Moon distance.
Material leaving the Earth with a velocity greater than the escape
velocity will remain in an
Earth-like orbit as it travels around the Sun (i.e. roughly circular with a 
semi-major axis of about 1 Astronomical Unit (AU)).  Over time, this material
will interact with the Earth-Moon system.  The goal of the numerical simulations is to determine the likelihood of impact
with the Moon and Earth during a relatively short period of time.

The dynamical studies in this paper describe the ejecta as a spherically 
symmetric distribution traveling radially away from the planet, similar 
to the assumption adopted by Gladman 1997.  In order to account for a range of velocities,
we simulate the 
particles with velocities, $v_{\infty}$, from $0.0$ to $3.3 \ km \ s^{-1}$ , which is
related to the launch speed, $v_l$, by 

\begin{equation}
{v_{l}^{2}} = { v_{esc}^{2}} + {v_{\infty}^{2}}.
\end{equation}

\noindent Since the Earth's escape velocity is large, this corresponds to a relatively narrow range of velocities, from $11.2 \ km \ s^{-1}$ to roughly $11.7 \ km \ s^{-1}$.

Each simulation 
consists of 225 ejecta particles, plus the nine planets and the Moon.  The 
simulations were integrated using the dynamical integration code {\it pkdgrav}
(Stadel 2001), employing 
a variable timestep to resolve close interactions with planets and resolve
any 
collisions.  Since the simulation 
includes a large range of orbital timescales the maximum timestep is determined
by the object with the shortest orbital period which in our case is the Moon. To insure
detailed 
calculation of the particle orbits, the number of integrations is 
maximized to a time resolution
of 200 time steps per orbit.  
Given that the Moon is included and has the shortest
dynamical time (orbiting the Earth once every 5.9 days at a distance of 
21.6 $R_\oplus$), the maximum timestep is
limited to 42 minutes.  The minimum timestep is 1.5 minutes, sufficient to 
resolve any close encounters and collisions.  The initial positions,
velocities, and masses of the nine planets were taken from the DE 403
ephemeris, provided by the Jet Propulsion Laboratory.
All of the simulations were
run for 4775 years, each representing about 1 week of computing time on a
modern
desktop machine.
The simulation results are recorded at approximately 5 year 
intervals for the entire integration.   

During potential particle-planet encounters,
the timestep is reduced to allow an accurate calculation of the
interaction. The code uses a 16 rung multistepping ladder to reduce the time step from the maximum to the minimum required to resolve the encounter. 
Collisions are resolved at a timestep, $n$, by linearly extrapolating to possible encounters at the next timestep, $n+1$, using velocities and positions of the particles at timestep $n$.  Therefore, collisions with a small particle and a planet are well resolved, given a small enough timestep.  However, errors in impact velocities can be significant.  The velocity error is equal to $\delta T \ a$, where $\delta T$ is the timestep, and $a$ is the acceleration.  At 1 AU from the Sun, in  the absence of other potentials, this velocity error is only $1.5 \ m \ s^{-1}$.  However, at the Earth's surface, it can be as great at $800 \ m \ s^{-1}$.  To test the robustness of the code, we computed one simulation with half the timestep to make sure we recorded the same number of collisions.

For the purposes of this study, we are interested in orbital transfer of 
material over a relatively short period of time, less than 5000 years.  
We chose this interval to bias our results toward material that would be rapidly buried on the lunar surface, thus protecting it from degradation by UV radiation, cosmic rays, and micrometeorite weathering.  However, since the material will continue to be transferred even beyond 5000 years, our results must be considered a lower limit on the total amount of Terran material delivered to the Moon by this process.

{\bf [ Figure 1 ]}

Looking more closely at the re-accretion of material onto the Earth,
Fig. 1 shows the fate of 138 impacts in the $v_\infty = 0.0 \ km \ s^{-1}$
case, depicting the cumulative number of impacts as a function of time.
We see that $89 \%$ of re-accretion collisions occur within the first 
100 years of the simulation, and $95 \%$ of the collisions have occurred
within 1000 years.  The remaining 5\% of the reaccretion occurs between 1000 and 5000 years.
We see from Table 2 that the quantity of ejecta returning to the Earth is 
greatest for $v_{\infty} = 0.0 \ km \ s^{-1}$
 and rapidly decreases for increasing
$v_{\infty}$. Over such a short period of time, and with simulations 
containing such small numbers of particles,  lunar interactions with the 
ejecta are rare.  We therefore estimate the amount  of material interacting 
with the Moon from simple scaling arguments.  We expect the ratio of the number 
of Earth impacts, $N_{\oplus}$, to the number of lunar impacts, $N_{Moon}$ to be

\begin{equation}
\frac{N_\oplus}{N_{Moon}} = \frac{R_{\oplus}^2}{R_{Moon}^2} 
		\frac{v_\infty^2 + v_{esc,\oplus}^2}{v_\infty^2 +
	         v_{esc,m}^2 + v_{pot}^2},
\end{equation}

\noindent where $v_{esc,\oplus}$ and $v_{esc,m}$ are the escape
velocities of the Earth and Moon, and $v_{pot} = \frac{v_{esc,\oplus}}{\sqrt{\eta}}$ represents the contribution of the Moon's proximity to
the Earth's potential well.  For the distance to the Moon chosen for our study,
21.6 $R_\oplus$, $v_{pot}$ is approximately equal to the Moon's escape velocity,
and gives a derived ratio of 140 for $v_\infty = 0.0 \ km \ s^{-1}$.  
This matches well with the 
simulation for $v_\infty = 0.0 \ km \ s^{-1}$,
 and allows us to scale the other results 
accordingly.  Table 2 indicates these scaled results, along with derived
lunar impact fractions and transfer efficiencies, $F_e$.  We use Eq. 4 to determine the amount of material launched within a given velocity range, and use these values to derive the fraction of impactor mass transferred to the Moon, $F_{orb}$.
These will be used in the following section to derive the mass flux 
from Earth to the Moon.

{\bf [ Table 2 ]}

\subsection{{\small {\bf Lucky shots}}}

Finally, we take into account the  ``lucky shots'' that impact
the Moon on their way out of the Earth-Moon system.  To first order,
for fast moving particles, the fraction hitting the Moon is just the 
cross sectional area of the Moon divided by the surface area of an
imaginary sphere encompassing the lunar orbit:

\begin{equation}
F_{lucky} = \frac{R_{Moon}^2}{4 D_{Moon}^2}.
\end{equation}

\noindent For the same conditions as the numerical simulation, namely $D_{Moon}
= 21.6 \ R_{\oplus}$, $F_{lucky} =
4.0\times10^{-5}$, the same as the transfer efficiency for the
$v_{\infty} = 3.3 \ km \ s^{-1}$ case. Given this similarity,
when these high velocity ``lucky shots''  are
included in the mass flux calculation, they are treated with the
high velocity particles from the numerical simulations. 

\subsection{{\small {\bf Lunar impact velocities}}}

The sections above determine the fraction of ejecta reaching the Moon.  However, the velocity distribution of this material must be calculated to determine the likely impact velocities on the  Moon expected during lunar encounters.  Since lightly shocked millimeter-sized chondritic meteorites are found in the lunar soil
(McSween 1976; Melosh 1989), there is some hope of intact sample
survival of even fast moving impactors.

To explore the impact velocities, we first address the velocity distribution of the ejecta launched from the surface of the Earth.  Eq. 4, above, from Chyba {\em et al.} 1994 gives the integrated probability distribution in terms of the impactor mass.  We start by taking the derivative of this equation to get the velocity distribution function for velocities within a range from $v$ to $v + \delta v$, 

\begin{equation}
P_{v} = C \frac{\partial M\left(>v\right)}{\partial v},
\end{equation}

\noindent where C is the normalization constant given by

\begin{equation}
C = \frac{1}{\int_{v_{min}}^{v_{max}} \frac{\partial M\left(>v\right)}{\partial v}dv}.
\end{equation}

\noindent We take $v_{min}$ and $v_{max}$ from the full range of launch velocities, from 10.94 to 11.7 $km \ s^{-1}$.  Thus, the probability of finding a launch velocity within the range of $v$ and $v + \delta v$ is $\int_{v}^{v + \delta v} P_{v} dv$.  Through conservation of energy in the Earth's potential and knowledge of the Moon's orbital velocity, $v_{orb}$, we can translate this into the interaction velocity at the Moon.  Added to this is the Moon's escape velocity, $v_{esc,m}$, to get the maximum impact velocity for a given launch velocity

\begin{equation}
v_{i, max} = \left(v_{int}^2 + v_{orb}^2 + v_{esc,m}^2\right)^{\frac{1}{2}},
\end{equation}

\noindent where $v_{int}$ is determined by

\begin{equation}
v_{int} = \left(v_l^2 + v_{esc, \oplus}^2\left(1 - \frac{1}{\eta}\right)\right)^\frac{1}{2}.
\end{equation}

\noindent The maximum impact velocity tells us about the total kinetic energy carried by the particle as it impacts the Moon's surface.  This corresponds to a velocity of around $5 \ km \ s^{-1}$ with $\eta = 21.6$.  However, the impactor will disperse some of this energy during oblique impacts, and therefore the maximum impact velocity doesn't necessarily determine the amount of shock delivered to the impactor.  Accordingly, we compute the vertical component of the velocity as the upper bound on the stress induced in the impactor.  Measuring the impact angle, $\theta$, from the ground (i.e. 90 degrees is a direct impact) the vertical component is given by

\begin{equation}
v_{vert} = v_{i,max} sin{\theta}.
\end{equation}

The probability of a given particle having an impact angle between $\theta$ and $\theta + \delta \theta$ is given by  $\int_{\theta}^{\theta + \delta \theta} P_{\theta} d\theta$.  Pierazzo and Melosh 2000a show that this normalized distribution is

\begin{equation}
P_{\theta} = 2 sin{\theta}cos{\theta}
\end{equation}

\noindent regardless of the planet's gravity field.  Using this information, along with our velocity distribution, we compute the expected value of the vertical impact velocity, $v_{exp}$, over our range of velocities and for impact angles less than or equal to $\theta$

\begin{equation}
v_{exp}\left(\theta\right) = \int_{v_{min}}^{v_{max}} \int_{0}^{\theta}P_{\theta}P_{v}v_{vert}d\theta dv.
\end{equation}

\noindent Table 3 gives $v_{exp}$ for a range of impact angles, and the corresponding probabilities for such angles, determined from Eq. 15. 

{\bf [ Table 3 ]}

We can use the results from the orbital simulations to spot check the lunar impact calculations.  We have one impact each for the $v_{\infty} = 0.0 \ km \ s^{-1}$ and $v_{\infty} = 1.8 \ km \ s^{-1}$:  one at $3.9 \ km \ s^{-1}$ and one at $4.2 \ km \ s^{-1}$ occurring 140 years and 541 minutes after launch, respectively (This last impact represents one of the "lucky shots" impacting the Moon as it leaves the system.)  These impact velocities correspond to the respective $v_{i,max}$ for each case.  Both of the impacts occurred at roughly 30 degree impact angles, corresponding to a vertical velocity of $2.0 \ km \ s^{-1}$ and $2.1 \ km \ s^{-1}$, well within our most probable calculated vertical impact velocity.

The impact velocities also give us a rough estimate of the likelihood that an aggregate chunk of material will survive the impact in any recognizable form.  Given the nature of our calculations, we want to derive the impact pressure experienced by the rock fragment upon impact.  To get an order of magnitude estimate, we use simple scaling and dimensional analysis to estimate this quantity.

The total deceleration experienced by the rock fragment upon impact is approximately

\begin{equation}
a = \frac{v_{vert}}{\Delta T},
\end{equation} 

\noindent where $v_{vert}$ is the vertical velocity and $\Delta T = \frac{d}{v_{vert}}$ is the time it takes to traverse the diameter, $d$, of the particle.  We can estimate the mean impact pressure experienced by a particle by multiplying this deceleration by the mass of the particle and dividing by the cross sectional area of the particle.  Thus, we derive an order of magnitude estimate for the mean impact pressure which is independent of the size of the impactor,

\begin{equation}
P_{ave} = \frac{2}{3}\rho v_{vert}^2,
\end{equation} 

\noindent with $\rho = 2860 \ kg \ m^{-3}$ for basalt.  Table 3 shows the results for a range of likely impactor velocities.  These values are in rough agreement with the literature-based constraints we impose for ejecta leaving the surface of the Earth as described in Section 4.  It should also be noted that dissipative effects of the regolith and other factors may reduce the impact pressure.  Thus, for the range of impact velocities calculated in this section, the likelihood of Terran ejecta surviving in some large aggregate sample are quite high, as will be discussed in detail later.

\section{{\small {\bf MASS TRANSFER RATES}}}

\subsection{{\small {\bf Mass transfer from the Earth to the Moon}}}

Armed with transfer efficiencies from our analytical 
calculations and numerical simulations,
we can compute the mass flux of Terran ejecta incident on the Moon.
We will proceed in four steps:

\begin{enumerate}
\item{Calculate the mass of material 
incident on the Earth during the period of interest, from 3.9 Ga to the present, scaled from the
lunar impact record.}
\item{Determine the fraction of material that leaves the Earth 
with a given velocity during an impact event.}
\item{Apply the transfer efficiencies determined in the 
previous section to estimate the mass of material that reaches the
Moon.  Ejecta with velocities between $10.9 \ km \
s^{-1}$ and $11.2 \ km \ s^{-1}$ correspond to direct transfer efficiencies.
Ejecta with larger velocities correspond to either the orbital
transfer or lucky shot efficiencies.}
\item{Determine the fractional volume of this material in
the lunar regolith by considering the other material accreted by the Moon 
at the
same time, namely micrometeorite flux and ejecta from subsequent 
lunar impacts.}
\end{enumerate}

To gain an estimate of the mass flux incident on the Earth as a function of
time, we follow the method described in Chyba {\em et al.} 1994, making use of the lunar cratering record.
   The cumulative crater density for craters greater than a
given size, $D$, due to impactors incident on the Moon as a function of 
time can be modeled by an equation of the form

\begin{equation}
N\left( t, >D \right) = f\left( t \right)\times
	 \left( \frac{D}{4\  km} \right) ^{-1.8} km^{-2}
          \ Gyr^{-1}.
\end{equation}

\noindent We consider two possible variations in the time dependence
of this model.  First, we explore the model
employed by Chyba {\em et al.} 1994 which describes the impact history as being roughly 
constant for the past 3.5 Ga, and increasing exponentially for earlier times,

\begin{equation}
f_{exp}\left( t \right) =  \alpha \left( t + \beta e^{t/\tau} \right).
\end{equation}

\noindent We also consider the case of the lunar cataclysm, which includes a 
period of increased impact flux about 3.9 Ga.  This is modeled by adding 
a Gaussian term,

\begin{equation}
f_{cat}\left( t \right) =  f_{exp}\left( t \right) + \gamma e^\frac{-\left( t - \mu \right)}{2\tau_{c}^2}.
\end{equation}

\noindent with $\mu = 3.9$ Ga, the 
peak of the cataclysm, and $\tau_c$ is the gaussian width of the event.
Each of these equations are fit to the lunar data provided in
the BVSP 1981.  
Since the lunar data are cumulative in both crater size and time, 
the data must be fit by the cumulative number density distribution

\begin{equation}
N_c \left( 4 \ km \right) = \int_{0}^{t_{age}} 
	N\left( t, 4 \ km \right) \ dt \ km^{-2},
\end{equation} 
 
\noindent with $D = 4 \ km$ as shown in Fig. 2.  
Similar to Chyba {\em et al.}, we perform a 
$\chi^2$ minimization with the coefficients as free parameters,
 leaving the decay time,
$\tau$, and the width of the cataclysm, $\tau_c$, fixed for each case (as noted in the table, $\tau$ for the case of the lunar cataclysm was modified slightly from Chyba's model to improve the fit).  Table 4 details the 
parameter values and fit results graphically depicted in Fig. 2.  The 
values for $\alpha$ and $\beta$ given in Table 4 for the exponential case
differ from those determined by Chyba {\em et al.} 1994, but the differences are 
not significant and can be traced to differences in the interpretation of
data from BVSP.

{\bf [ Table 4 ]}

{\bf [ Figure 2 ]}

Following arguments similar to Chyba {\em et al.} 1994, we now calculate the mass 
incident on the Moon as a function of time.  The mass of an 
impactor can be related to the final crater diameter,

\begin{equation}
m \left( D \right) = 0.54 \Gamma v^{-1.67} D_c^{0.44} D^{3.36}
\end{equation}

\noindent where $\Gamma = 1.6 \times 10^{3}$ (with units of $kg \ s^{-1.67}$)
is a constant for lunar values of the surface gravity and surface density 
(Chyba {\em et al.} 1994), $v$ is the impactor velocity in $m \ s^{-1}$, 
$D_c = 11,000$ is the transition from craters with simple bowl shapes to more complex 
morphologies, and $D$ is the final crater diameter, both in meters.  
Combining Eq. 23 with Eq. 19 gives 
a cumulative crater density for objects greater than a given mass, $m$,

\begin{equation}
n\left( t, > m \right) = f\left( t \right)\times
	 \left( \frac{m}{ m \left( 4\  km \right)} \right) ^{-b} km^{-2}
          \ Gyr^{-1}.
\end{equation} 
 
\noindent where $ b = \left( 1.8/3.36 \right) = 0.54$.  Furthermore, 
we can determine the total mass of impactors incident on the lunar surface
per unit area as a function of time by integrating over the mass 
distribution

\begin{equation}
 M \left( t \right) = \int_{m_{max}}^{m_{min}} m  
	\frac{\partial n}{\partial m} dm.
\end{equation}

\noindent Upon integration and the assumption that $m_{max} \gg m_{min}$,
the total mass of objects incident on the surface per $km^{2}$ per Gyr 
is given by

\begin{equation}
 M \left( t \right) =  f \left( t \right) m(4 \ km)^b \frac{b}{1-b} 
			m\left( D_f \right)^{1 - b}, 
\end{equation} 

\noindent where $m \left( D_f \right)$ is the maximum mass impactor
represented by the largest impact basin, taken by Chyba to be 
the $D_f = 2200 \ km$ South Pole-Aitken crater. 

In scaling these impactor fluxes to the Earth, we take into account the
increased gravitational focusing of the Earth through the ratio
of impact cross sections,

\begin{equation}
\frac{\sigma_{\oplus}}{\sigma_{Moon}} = \frac{R_{\oplus}^2}{{R_{Moon}}^2}
	\frac{1 + \left( \frac{v_{esc,\oplus}}{v_i} \right)^2}
	{1 + \left( \frac{v_{esc,m}}{v_i} \right)^2},
\end{equation}

\noindent where $R_\oplus$ and $R_{Moon}$ are the radii of the Earth and Moon,
$v_{esc,\oplus}$ and $v_{esc,m}$ are the escape velocities of the Earth and 
Moon, and $v_i$ is the typical velocity of the interacting particle.
If the particles are asteroids, with typical impact speeds of 
$14 \ km \ s^{-1}$ (Chyba {\em et al.} 1994; Bottke {\em et al.} 1994), this ratio is $\sim$ 24.
Multiplying $M\left( t \right)$ by the area of the Moon and 
taking into account the ratio of impact cross sections gives the total
mass of material incident on the Earth as a function of time.  

Finally, we consider the transfer of material from the Earth to the Moon.
In Section 2, we computed the total fraction of material transfered directly from the Earth to the Moon, $F_{direct}\left(t\right)$, as a function of time.  The mass transferred via this method is 

\begin{equation}
M_{dir}\left(t\right) = F_{direct}\left(t\right)M\left(t\right)
\end{equation}

For orbital transfer, we determined the fraction of the impactor transferred
from the Earth to the Moon, $F_{orb}$, at a distance of 21.6 $R_\oplus$.
The mass of material ejected from the Earth at speeds greater than the
escape velocity during these impacts is proportional
to the mass of the incident material, $m$, determined from Eq. 4.
Using  $\rho = \rho_t = 2860 \ kg \ m^{-3}$, we derive that $M_e \left( > v_{esc}
\right) = 0.14 \ m$.

Since the Earth does the bulk of the gravitational focusing when interacting with Sun-orbiting particles, as the Moon recedes from the Earth, the transfer efficiency
will decrease in proportion to the ratio of gravitational potential
of the Earth at a distance $R_0 = 21.6 R_\oplus$ and the 
instantaneous value, $D\left( t \right)$,

\begin{equation}
F_{orb}\left( t \right) = F_{orb} \frac{R_0}{D \left( t \right)}.
\end{equation}

\noindent We take the instantaneous value of $D\left( t \right)$ to
have the functional form described by Williams 2000,

\begin{equation}
D\left( t \right) = R_0 \left[ 1 - \frac{13}{2} 
	\frac{\langle \dot{R} \rangle}{R_0}
	\left( T_{ref} - t \right) \right]^{2/13}, 
\end{equation}

\noindent where $\langle \dot{R} \rangle = -673 \ R_\oplus \ Gyr^{-1}$ 
is the average lunar recession velocity, and $T_{ref}$ = 3.9 Gyrs 
is the reference time for our
simulations when the Moon was 21.6 $R_\oplus$ from the Earth.  
$\langle \dot{R} \rangle$ is determined by taking $D\left( 0 \right)
= 60.3 \ R_\oplus$, the current lunar distance. 

To determine the total mass transferred to the Moon, we multiply the ejected material by the transfer fraction

\begin{equation}
M_{orb}\left( t \right) = F_{orb}\left(t\right) M\left( t\right).
\end{equation}

\noindent Table 2 shows the fraction of material launched for a given $v_{\infty}$.
The ``lucky shots'' are included in the last integral, since the
transfer efficiencies are the same as for $v_{\infty} = 3.3 \ km \
s^{-1}$.  The relative contributions from orbital and direct transfer show that orbital transfer dominates for the last 3.9 Ga.  At 3.9 Ga, orbital transfer accounts for 58\% of the total transferred mass.  The orbital contribution almost completely dominates at the present epoch.

It should be noted that these transfer methods ignore the inhibiting effects produced by the Earth's oceans and the Earth's atmosphere.  However, we are considering only large impactors, and our condition that $m_{max} \gg m_{min}$ must be satisfied.  To put this in perspective, the mass of the impactor that formed the South Pole-Aitken Crater is roughly $1.5 \times 10^{19} \ kg$.  If we only consider impactors larger than $10 \ km$ (roughly the scale height of the Earth's atmosphere, and also larger than the depth of the Earth's ocean), this corresponds to a mass of $1.5 \times 10^{15} \ kg$, a factor of $10^{-4}$ less than $m_{max}$.  In addition, impacts of this size should excavate enough atmosphere for ejected material to leave the Earth with minimal interaction (Melosh 1988).

\subsection{{\small {\bf Burial on the lunar surface}}}

As the Terran material reaches the Moon, it will be buried by 
the accretion of micrometeorites and ejecta from subsequent lunar impacts.
Love and Brownlee 1993 measured directly the flux of micrometeorites
incident on the Earth today to be $M_{micro} = 4 \times 10^{16} 
\ kg \ Gyr^{-1}$.
In the absence of other constraining evidence, we take this to follow
the time dependence of Chyba {\em et al.} 1994, and scale it
to the Moon by considering the difference in surface area and gravitational
focusing between the Earth and Moon.

The mass of ejecta, crudely averaged over the
surface of the Moon, can be determined from methods described in Sleep
and Zahnle 2001.
We scale the mass of ejecta produced in a given impact to the volume
of material removed from the crater.  Cast in terms of the Moon's
escape velocity and the mass of the impactor, the relationship is

\begin{equation}
m_{ej}\left(m\right) = 18.2
\left(R_{Moon}\right)^{0.65}\left(\rho_{t}\right)^{0.217}\left(\frac{v_{i}}{v_{esc,m}}\right)^{1.3}m^{0.783},
\end{equation}

\noindent where $v_{i}$ is the impact velocity on the Moon, $R_{Moon}$
is the radius of the Moon in meters, and $\rho_{t}$ is the density of
the lunar surface, taken to be $2860 \ kg \ m^{-3}$, and $m$ is the
impactor mass, in kilograms.  To
determine the total amount of ejecta produced on the Moon as a
function of time, we integrate this over the impactor distribution,
Eq. 25,

\begin{equation}
M_{ej}\left(t\right) = \int_{m_{max}}^{m_{min}}m_{ej}\left(t\right)
\frac{\partial n}{\partial m} dm.
\end{equation}

\noindent Again, assuming $m_{max} \gg m_{min}$ and $b = 0.54$, the total
mass of ejecta as a function of time is

\begin{equation}
M_{ej}\left(t\right) = 40.4
\left(R_{Moon}\right)^{0.65}\left(\rho_{t}\right)^{0.217}\left(\frac{v_{esc,m}}{v_{i}}\right)^{1.3}
m\left(4 \ km\right)^{b}f\left(t\right)m\left(D_{f}\right)^{0.243}.
\end{equation}

\noindent Thus, the total amount of ejecta available to bury Terran
material is this amount minus the amount lost to space, determined by
Eq. (4), with appropriate lunar parameters. 

At this point, for each of the contributions of Terran sources, 
micrometeorites, and subsequent lunar ejecta (collectively denoted $M_{cont}$, below),
we define an equivalent thickness of material per square kilometer per Gyr (averaged 
over the surface of the Moon)

\begin{equation} 
H\left( t \right) \equiv \frac{M_{cont}\left( t \right)}{\rho},
\end{equation}

\noindent with the densities, $\rho$, of the various materials taken 
to be $2860 \ kg \ m^{-3}$ for the ejecta and Terran material, 
and $2200 \ kg \ m^{-3}$
for the micrometeorites.  The fractional volume of Terran material 
accreted subsequent to a time, $t_{age}$, is determined by integrating 
each of the contributions up to the age of the surface in question and 
taking the ratio of the volume of Terran material to the other sources

\begin{equation}
\frac{V_{ter}}{V_{other}} = \frac{\int_{0}^{t_{age}}H_{ter}dt}
	{\int_{0}^{t_{age}}H_{micro} + 
	 \int_{0}^{t_{age}}H_{ej}dt}.
\end{equation}

\noindent Finally, the total depth of material accreted since a time
$t_{age}$ is determined by summing the contributions of all three materials,

\begin{equation}
H_{total} = \int_{0}^{t_{age}}H_{micro} + H_{ej} + H_{ter} \ dt.
\end{equation}	

Fig. 3 shows $H_{total}$ as a function of time, with most of the
material contributing to the depth of lunar regolith deposited during and 
shortly after 
the Late Heavy Bombardment, 
from 3.9 to 3.8 Ga.  Fig. 4 shows the fractional volume, or mixing
ratio, of Terran material as a function of time.  Again, the highest 
fraction occurs from 3.8 to 3.9 Ga, when the Moon's proximity to the
Earth facilitated transfer.  

{\bf [ Figure 3, Figure 4 ]}

These calculations, however, are a lower limit on the amount of Terran 
material accreted by the Moon.  Sleep {\em et al.} 1989, and subsequently Chyba {\em et al.} 
1994, point out that, due to small number statistics, the Moon
under-samples the largest impactors.  Chyba {\em et al.} 1994 predict more
impacts on the Earth, and argue that an impactor of mass $\sim 1.4 \times 10^{22} \ kg$ was probable subsequent to 4.4 Ga.
Using Eq. 4 and a simple average of our transfer efficiencies, $1.5 \times 10^{-6} $ times the mass of the impactor, as much 
as $ 6 \times 10^{8} \ kg \ km^{-2}$ 
of material will reach the Moon, corresponding to 
an equivalent basalt thickness of $0.2 \ km$.  However, it is 
impossible to tell when this material may have arrived on the Moon, 
depending as it does on a very few large impacts occurring 
stochastically through time.  In addition, these impactors were likely 
accreted long before the end of the Late Heavy Bombardment, and thus before 
the formation of the lunar maria.  Therefore, we ignore this contribution
in the above calculation, but note that this
very significant amount will most likely be buried and possibly protected by the formation of the 
maria.  However, if subsequent impacts break through the maria, this reservoir of Earth material may be exposed to the surface.  According to our calculations, it would be necessary to excavate to a depth of at least 300 meters (and most likely more) to gain access to this layer of Terran material.  Assuming a crater diameter to excavation depth ratio of 10 to 1 (Melosh, 1989), any crater 3 km in size or greater should provide a window to this layer.

Finally, the effects of gardening (the breakup and mixing of the lunar regolith
by micrometeorites and larger impactors) on the vertical distribution of this 
material will be substantial.  Lunar cores from the Apollo missions 
directly indicate vertical mixing of up to one meter (Mustard 1997),
and samples indicating deep mixing to tens of kilometers have been
found (Lindstrom and Lindstrom 1986).  While most of the very deep
mixing occured before the period of interest, we can expect that the
regolith will be well mixed to the depths indicated in this study.  Therefore, 
any sampling of the lunar stratigraphy will have a coarse time resolution
at best. This mixing may facilitate easier retrieval of samples over a range of potential ages, as subsequent small impacts will excavate older material directly to the surface.
The amount of Terran material on the surface of the Moon will depend
largely on the age of the surface that is searched.  
Assuming the regolith is well mixed, we estimate the total surface abundance of Terran to lunar material to be 7 ppm.  
This corresponds to $\sim 20,000 \ kg$ of Terran material over a 10 x 10 square km area..

\subsection{{\small {\bf Mass transfer from Venus and Mars to the Moon}}}

We also explored the fate of material
transferred from Venus and Mars to the Moon.  Due to Venus' proximity to the
Sun, the velocity required to launch a particle from the surface 
with enough energy to reach an Earth crossing orbit is $v_\infty = 3.3
\ km \ s^{-1}$.  We performed one simulation for Venus, tracking 225 particles
for about 5000 years.  After this period of time, only one particle 
hit the Earth.  Thus, subject to the errors of small number statistics,
the Venus-Earth transfer efficiency is $\sim 4 \times
10^{-3}$.  Using our scaling for the gravitational cross sections
of the Earth and Moon further reduces this transfer efficiency, on average, by another
factor of $\sim 10^{-5}$, giving a Venus-Moon transfer efficiency of 
$1.2 \times 10^{-7}$.  This is 30,000 times less than the Earth-Moon
transfer efficiency of $4 \times 10^{-3}$ for particles leaving the
Earth with the escape speed, indicating that Venus rocks on the
Moon will be a rare find indeed.  Still, an
area of 10 x 10 square km should still
yield almost 1 kg of Venusian material, if it can be identified as such.  Moreover, our estimate is a strict lower limit.  Melosh and Tonks 1993, in a simulation of the fate of ejecta from other planets, indicates that about 30\% of the ejecta leaving Venus with a wide range of velocities from $v_\infty = 0.0$ to $5 \ km \ s^{-1}$ hits the Earth within 12 million years.  This results in a Venus-Moon transfer efficiency of $\sim 3 \times 10^{-6}$, only about 1000 times less than the Earth-Moon transfer efficiency for particles launched from Earth with the escape speed.  This means up to 30 kg of Venusian material could exist in the same 10 x 10 square km area.

The situation for Mars rocks is more optimistic.  Estimates from
Gladman 1997 and Halliday {\em et al.} 1989 indicate that fifteen 100-gram
Mars rocks impact the Earth each year.  Scaling this to the Moon,
and assuming the amount of Martian material scales with the same time
dependence as the rest of the incoming debris, we expect $6 \times
10^{-8}$ of the lunar material to be Martian in origin, corresponding to
about 180 kg in the same 10 x 10 square km area.  
 
\section{{\small {\bf SURVIVABILITY OF BIOLOGICAL AND GEOCHEMICAL TRACERS}}}

Having calculated the mass fraction of Terran material on the Moon, we can now estimate the forms of evidence such material may contain. First, we identify physical constraints for specific evidence types. Then, we determine the planetary ejection and lunar impact regimes in which those constraints are satisfied. Finally, we convert this into a mass fraction of Terran material on the Moon corresponding to the given evidence type for different impactor velocities.

\subsection{{\small {\bf Constraints on survival of biological and geochemical evidence}}}

The pressures and temperatures experienced by the Terran material during ejection from Earth and deposition on the Moon will determine the survival of biological and geochemical evidence.
Although the
impact temperatures, pressures and fates of targets have been extensively
considered, little information is available about these conditions in the
projectile during and subsequent to impact (Pierazzo and Melosh 2000b).
Given this scarcity, and to avoid the complications of various equations
of state and impact parameters, we establish general and approximate
criteria to estimate the survivability of four different types of evidence:
isotopes, significant volatile inventories, organic carbon, and
molecular fossils (biomarkers).  
We emphasize that these calculations are intended only as an order 
of magnitude approximation for the amount of material experiencing 
conditions conducive to the possible survival of these evidence types.

\subsubsection{{\small {\bf Isotopes}}}

Carbon isotope systematics have been interpreted as
evidence for life on Earth as early as 3.85 Ga (Mojzsis {\em et al.} 1996).
Similarly, other isotopes could also serve as biosignatures, for example, nitrogen or oxygen (Blake {\em et al.} 2001).
Analysis of
biological isotope fractionations requires comparison of a putative
biological fractionation to the background signature of the host rock.
Effectively, this imposes the condition that the Terran material avoid
significant melting. 
Following Melosh 1988 and Stoffler {\em et al.} 1991, we set
the onset of melting at a pressure of 70 GPa, recognizing that this value
actually varies according to specific mineral assemblages and porosities.
We assume that most material in a $5 \ km \ s^{-1}$ impact on the Moon will not attain peak pressures greater than this and thus should avoid substantial melting, as indicated by Ahrens and O'Keefe 1971, Kipp and Grady 1996, and our own calculations (Table 3). Since all lunar impacts due to
direct and orbital transfer occur at velocities $< \ 5 \ km \ s^{-1}$, 
the fraction of
material remaining solid is determined by the material ejected from Earth
at pressures less than 70 GPa.

\subsubsection{{\small {\bf Significant volatile inventories}}}

Tyburczy {\em et al.} 1986
empirically determined the percent volatile loss of a carbonaceous
chondrite as a function of the projectile velocity. At velocities slightly
greater than $2.0 \ km \ s^{-1}$ and peak shock pressures 
$< 30$ GPa, 50 \%
of projectile volatiles were lost. We therefore limit ejection to 30 GPa
and lunar impact velocities $\leq 2.0 \ km \ s^{-1}$.

\subsubsection{{\small {\bf Organic carbon}}}

 Empirical (Tingle {\em et al.} 1992; Tingle 1998; Tyburczy {\em et al.} 1986)
and theoretical (Pierazzo and Chyba 1999) work suggests significant
organic carbon fractions can survive impacts. In the experiments of Tingle
{\em et al.} 1992, only small amounts of organic carbon were lost for impacts
up to $2 \ km \ s^{-1}$ and pressures up to 10 GPa. 
Although other work
has suggested more significant losses (e.g. Peterson {\em et al.} 1997), we
adopt as our constraints a maximum pressure during ejection of 10 GPa and
lunar impact velocities $\leq 2.0 \ km \ s^{-1}$.  We also note that the very fact that organic carbon can survive some impact conditions implies the need for rigorous criteria to distinguish extraterrestrial and terrestrial organic carbon.

\subsubsection{{\small {\bf Molecular Fossils}}}

To our knowledge, no information is available concerning survival of molecular fossils under impact conditions. In fact, impact conditions are likely to far exceed the nonetheless severe exigencies (temperatures $\sim$ 300 C, pressures $\sim$ 1 GPa) of prehnite-pumpellyite facies metamorphism, from which 2.7 Ga biological lipids have been successfully extracted (Brocks {\em et al.} 1999). Given the enormous informative value of biomarkers and the long-term stability of hydrocarbons (Mango 1991; Dutkiewicz {\em et al.} 1998), we believe it is not unreasonable to expect that new biomarkers, corresponding to the harsh conditions of lunar impact, could usefully be defined. Towards that end, we calculate the fraction of Terran material ejected from Earth at pressures $\leq$ 1 GPa and impacting the Moon at velocities $\leq 0.5 \ km \ s^{-1}$. 

\subsection{{\small {\bf Calculation of survival fractions}}}

	Until recently, the prospect that material could escape a planet via a natural process was considered extremely unlikely, much less that the material could do so without being heavily shocked (Melosh 1993). Experimental (Gratz {\em et al.} 1993) and observational evidence has forced a revision of this opinion. Most significantly, the SNC meteorites experienced shocks no greater than ~45 GPa, with corresponding temperatures $\le 600$ C (Stoffler 2000, as cited by Mastrapa {\em et al.} 2001). In fact, ALH84001 apparently traveled from the surface of Mars to Earth without ever exceeding 40 C (Kirschvink {\em et al.} 1997; Valley {\em et al.} 1997; Weiss {\em et al.} 2000). Two mechanisms might account for these observations. During an oblique impact, a vapor plume jet could entrain and accelerate surface material to the escape velocity with little shocking (O'Keefe and Ahrens 1986). However, at low impact angle, most or all of this material will be derived from the impactor, while the viability of this method at higher impact angles ($> 25$ degrees) is uncertain (Pierazzo and Melosh 2000b). A second and more probable method, spallation, results in lightly shocked material due to the interference of the stress and rarefaction waves near a free surface. Our calculations are limited to this case.

	The mass of spalled ejecta can be estimated using the method described in Melosh 1985,

\begin{equation}
M_e \left( > v_{e}\right) = \frac{0.75 P_{max}}{\rho_t c_L v_i} \left[\left(\frac{ v_i}{2 v_e}\right)^{\frac{5}{3}} - 1 \right] m,
\end{equation}

\noindent where $M_e$ is the mass of the ejecta, $v_e$ is the ejecta velocity, $P_{max}$ is the
maximum pressure experienced by this fraction of the ejecta, $v_i$ is the
velocity of the impactor, $\rho_t$ is the target density ($2860 \ kg \ m^{-3}$ for
basalt), $c_L$ is the longitudinal speed of sound ($6000 \ m \ s^{-1}$) , and $m$ is the mass of the
impactor.  This equation differs from that of Melosh 1985 only due to a mathematical 
error in the latter (for discussion and derivation of Equation 38, see 
Appendix 1).  

	Implicit to Eq. 38 is the assumption that the maximum ejecta 
velocity is one-half the impactor's. By making this conservative 
assumption, we preclude the escape of spalled ejecta for a most probable 
impactor velocity of $14 \ km \ s^{-1}$; indeed, we require more rare impacts at 
velocities exceeding $22.4 \ km \ s^{-1}$ for any spall escape.  This suggests that 
most (but not all) Terran material on the Moon will be heavily shocked. 
Such material would nonetheless yield significant scientific returns, 
for example, by dating and characterizing the Late Heavy Bombardment on early Earth 
and in the young Solar System. Moreover, uncertainties pertaining to 
the velocity distribution of early Earth impactors and to maximal ejecta 
velocities have been treated conservatively in this paper. Faster 
impactors may have been more abundant 4 Ga than now (Sleep {\em et al.} 1989).
Similarly, some laboratory experiments 
suggest ejecta velocities can actually approach 85\% of the impactor 
velocity (Curran {\em et al.} 1977), which would enable escape for a $14 \ km \ s^{-1}$ 
impactor.

Making these conservative assumptions we determine the percent of the total Terran material on the Moon that corresponds to the requirements set for each of the four classes of biological and geological evidence.  Using Eq. 38, an ejecta velocity of $11.2 \ km \ s^{-1}$, and the pressure
and lunar impact velocity constraints imposed for the four material classes defined above, the mass of ejecta (in units of impactor mass) is derived over a range of impactor
velocities (Table 5).  We then divide this number by the fraction of ejected material escaping Earth determined from Eq. 4.  Once deposited on the Moon, the material delivered at the end of the Heavy Bombardment is expected to be buried to a depth of roughly 6 meters within 5 million years, making protection from space radiation a second order effect.  Our results, in terms of the percentage of Terran material on the Moon, are recorded in Table 6.  For example, if all the material on the Moon were delivered by an impactor with velocity of $v_i = 22.5 \ km \ s^{-1}$, over a10 x 10 square kilometer search area, the amounts of material having endured conditions permissive to the survival of molecular fossils, organics, volatiles, and isotopic signatures from early Earth are 0.4 kg, 9.2 kg, 28 kg, and nearly 80 kg, respectively.  

{\bf [ Table 5, Table 6 ]}

\section{{\small {\bf CONCLUSIONS}}}

In this paper, we have explored the Moon as an ideal location to search for remnants of the early solar system, particularly samples of Earth not currently available to researchers.  The Moon's proximity and relatively unaltered surface makes research missions viable, and the Terran samples on the surface are unavailable anywhere else in the solar system.  We have argued that Terran materials are abundant and near the surface, with a significant fraction retaining their geochemical and biological signatures for detailed analysis.  In addition, since the majority of Terran samples date from the end of the Late Heavy Bombardment, the samples in the lunar ``attic'' are a unique probe of the early conditions on Earth, and potentially contain clues to the earliest forms of life.

The Terran material is delivered in one of two ways, through direct transfer or orbital transfer.  We have found that orbital transfer of material is more efficient for the time since the end of the Late Heavy Bombardment, with direct transfer of material only comparable in early epochs. The amount of Terran material, 7 ppm, is sufficiently large to consider a search mission.

Before any such mission is attempted, the current stock of lunar material (approximately 400 kg worth) should be searched for Terran material.  Given a concentration of 7 ppm, there should be roughly 3 grams of Earth material in the current lunar samples.  Since lunar fines contain more than 10 million particles per gram, a technique of infrared spectroscopy coupled with microscopic imagery could distinguish hydrated silicates (common on Earth) from the dry, unhydrated lunar fines.  Such material could then be isotopically analyzed to confirm its terrestrial origin.  While this is not likely to yield much in the way of information about the early Earth, it would act as a proof of concept and a baseline for future missions.

Recovery and identification of Terran samples from the Moon represents a significant challenge.  Still, returning a sample from the Moon bearing the remains of Earth life may be orders of magnitude easier than returning, say, Martian samples from Mars bearing the remains of Martian life.  In this sense, lunar missions represent an interesting proving ground for these types of endeavors.  The risk of contamination and relative scarcity of Terran material makes sample return missions difficult.  Therefore, robotic missions need to be developed 
capable of finding Terran material using advanced in situ measuring 
devices to help identify samples and largely driving the analysis on the 
Moon.  The existence of a facility on the Moon to recover and analyze samples would guard against any contamination from Earth.  However, for a complete analysis of the material, we suggest the best way to conduct these studies is on site measurements by human observers - in essence, a return to the Moon.

\newpage

\begin{center}
{\bf ACKNOWLEDGEMENTS}
\end{center}
We thank Tom Quinn, Joachim Stadel, and Derek Richardson 
for providing the dynamics simulation
software, PKDGRAV, and other analysis and visualization tools used in the 
numerical simulations.  We also thank Don Brownlee, Toby Smith, and Katherine Shaw for 
discussions concerning the fate of Terran ejecta on the surface of the Moon.
Finally, we appreciate Chris Stawarz and Zoe Leindhardt for 
their assistance in the numerical modeling.  The paper benefited greatly from a thorough review by Brett Gladman, and additional reviews by Norm Sleep and Jay Melosh.  Also, thanks to Jay Melosh for helping confirm our new derivation of Eq. 38.  This research was supported by the NSF-IGERT trainingship in Astrobiology, an NDSEG fellowship, and the NASA Astrobiology Institute.

\appendix
\renewcommand{\thesection}{{\small {\bf Appendix \Alph{section}.}}}
\section{{\small {\bf Justification of Eq. 38}}}

The expression for the mass of spalled ejecta as a function of ejection velocity is based on that described and physically justified by Melosh 1985.  Briefly, Melosh derived an equation for the spall thickness, $z_s$: 

\begin{equation}
z_{s} = \frac{T D}{\rho_t C_L v_e}, 
\end{equation}

\noindent where $T$ is the dynamic tensile strength and $D$ is the impactor diameter.  Integration of this equation gives the volume of spalled ejecta.

An expression for $v_e$ can be deduced from the supplementary material to Melosh 1985, still available from GSA,

\begin{equation}
v_e = 2 v_i \left(\frac{D}{2 R}\right)^{3}, 
\end{equation}

\noindent where R is the radial distance from the impact epicenter to the point of ejection.

Thus, $z_s$ is inversely proportional to $v_e$ and directly proportional to the cube of $R$.  If a sign error is made, such that $z$ is mistakenly assumed proportional to $R^{-3}$, one derives the equation given in Melosh 1985. 

Avoiding that error, the volume of spalled ejecta can be determined by evaluating the integral 

\begin{equation}
\int_{R_{min}}^{R_{max}} 2 \pi R z_s\left(R\right) dR,
\end{equation}

\noindent where the limits of integration are defined by the maximum and minimum ejecta velocities being considered, corresponding to $R_{min} = \left(\frac{D}{2}\right)\left(4\right)^{\frac{1}{3}}$ and $R_{max} = \left(\frac{D}{2}\right)\left(\frac{2 v_i}{v_e}\right)^{\frac{1}{3}}$.  The maximum ejecta velocity, corresponding to $R_{min}$, has been assumed equal to half the impactor's velocity, as in Melosh 1985. 

Evaluating this integral and substituting $T = \frac{P_{max}}{\beta}$ (see footnote 4 in Melosh 1985), with $\beta \sim 4$, gives the equation used in this paper.

\bigskip
\noindent{\bf REFERENCES}
\bigskip 

\noindent Ahrens, T. and J. O'Keefe 1971. Shock melting and vaporization of lunar
rocks and minerals. {\it The Moon}, 4, 214-249.
\smallskip

\noindent Awramik, S. 1982. 
In {\it Mineral Deposits and the Evolution of the Biosphere} 
edited by H. D. Holland, Springer, NY, pp. 309-320.
\smallskip

\noindent Blake, R., J. Alt,  and A. Martini 2001. Oxygen isotope ratios of PO4: An
inorganic indicator of enzymatic activity and P metabolism and a new
biomarker in the search for life. {\it Proceedings of the National Academy of
Science USA}, 98(5), 2148-2153.
\smallskip

\noindent Bogard, D. and P. Johnson 1983. Martian gases in an Antarctic meteorite?
{\it Science}, 221, 651-654.
\smallskip

\noindent Bottke, W. F. Jr., M. C. Nolan, R. Greenberg, and R. A. Koolvoord 1994. Collisional lifetimes and impact statistics of near-Earth asteroids. In {\it Hazards due to Comets and Asteroids}, edited by T. Gehrels, University of Arizona Press, Tucson, pp. 337-357.
\smallskip

\noindent Brasier, M. D., O. R. Green, A. P. Jephcoat, A. K. Kleppe, M. J. Van 
Kranendonk, J. F. Lindsay, A. Steele and N. V. Gressineau 2002 .
Questioning the evidence for Earth's oldest fossils. {\it Nature}, 416, 76-81.
\smallskip

\noindent Brocks, J., G. Logan, R. Buick, and R. Summons 1999. Archean Molecular
Fossils and the Early Rise of Eukaryotes. {\it Science}, 285, 1033-1036.
\smallskip

\noindent Buick, R. 1990. Microfossil recognition in Archaean rocks: An appraisal of 
spheroids and filaments from a 3500 million year old chert-barite unit at 
North Pole Western Australia. {\it Palaios}, 5, 441-459.
\smallskip

\noindent BVSP, Project leaders: McGetchin, T. R., R. O. Pepin, 
R. J. Phillips 1981.  Basaltic volcanism on the terrestrial planets.
Pergamon Press, New York.
\smallskip

\noindent Chyba, C. F., T. C. Owen, W.-H. Ip 1994.
 Impact delivery of volatiles and organic molecules to Earth. 
 In Hazards Due to Comets and Asteroids, edited by T. Gehrels,
 University of Arizona Press, Tucson, pp. 9-58.
\smallskip

\noindent Curran, D., D. Shockey, L. Seaman and M. Austin 1977. Mechanisms and
models of cratering in earth media. In Impact and Explosion Cratering,
edited by D. J. Roddy, R. O. Pepin and R. B. Merrill, Pergamon Press, New
York, pp. 1057-1087.
\smallskip

\noindent Dickey, J.O. et al. 1994. Lunar 
Laser Ranging -- A Continuing Legacy of the Apollo Program. {\it Science}, 
265, 482.
\smallskip

\noindent Dutkiewicz, A., B. Rasmussen and R. Buick 1998.  Oil preserved in fluid
inclusions in Archaean sandstones. {\it Nature}, 395, 885-888.
\smallskip

\noindent Fedo, C. M. and M. J. Whitehouse. 2002. Metasomatic origin of 
quartz-pyroxene rock, Akilia, Greenland, and Implications for Earth's 
Earliest Life. {\it Science}, 296, 1448-1452.
\smallskip

\noindent Friedmann, E., J. Wierzchos, C. Ascaso, and M. Winklhofer 2001. Chains
of magnetite crystals in the meteorite ALH84001: Evidence of biological
origin. {\it Proceedings of the National Academy of Science USA}, 98(5),
2176-2181.
\smallskip

\noindent Gladman, B. 1997. Destination: Earth.  Martian meteorite delivery. {\it Icarus}, 130, 
228-246.
\smallskip

\noindent Gratz, A. J., W. J. Nellis, and N. A. Hinsey 1993. Observations of high-velocity, weakly shocked ejecta rom experimental impacts. {\it Nature} 363, 522-524.
\smallskip

\noindent Halliday, I., A. T. Blackwell, A. A. Griffin 1989. The flux of meteorites on the earth's
surface {\it Meteoritics} 24, 173-178.
\smallskip

\noindent Kipp, M. and D. Grady 1996. Experimental and numerical studies of
high-velocity impact fragmentation. In {\it High-Pressure Shock Compression of
Solids II: Dynamic Fracture and Fragmentation}, edited by L. Davison, D.
Grady and M. Shahinpoor, Springer-Verlag, New York, pp. 282-339.
\smallskip

\noindent Kirschvink, J. L., A. T. Maine, and H. Vali 1997. Paleomagnetic evidence of a low-temperature origin of carbonate in the Martian meteorite ALH84001. {\it Science} 275, 1629-1633.
\smallskip

\noindent Lindstrom, M. M., and D. J. Lindstrom 1986. Lunar granulites
and their precursor anorthositic norites of the early lunar
crust. {\it JGR}, 91, D263-D276.
\smallskip

\noindent Love, S. G., and D. E.  Brownlee 1993.
A Direct Measurement of the Terrestrial Mass Accretion Rate of Cosmic Dust.
{\it Science}, 262, 550.
\smallskip

\noindent Mango, F. 1991. The stability of hydrocarbons under the time-temperature conditions of petroleum genesis. {\it Nature} 352, 146-148.
\smallskip

\noindent Mastrapa, R. M. E., H. Glanzberg, J. N. Head, H. J. Melosh, and W. L. Nicholson
2001. Survival of bacteria exposed to extreme acceleration: Implications 
for panspermia. {\it Earth and Planetary Science Letters},  189, 1-8. 
\smallskip

\noindent McKay, D., E. Gibson, K. Thomas-Keprta, H. Vali, C. Romanek, S. Clemett, X. Chillier,  C. Maechling, and R. Zare 1996. Search for past life
on Mars: Possible relic biogenic activity in Martian meteorite ALH84001.
{\it Science}, 273, 924-930.
\smallskip

\noindent McKinnon, W. B., K. J. Zahnle, 
B. A. Ivanov, and H. J. Melosh 1997. Cratering on Venus: Models and Observations. 
in {\it Venus II}, edited by S. W. Bougher,
D. M. Houten, and R. J. Phillips, University of Arizona Press, Tuscon, pp.
969-1014.
\smallskip

\noindent McSween, H 1994. What we have learned about Mars from SNC meteorites. {\it Meteoritics} 29, 757-779.
\smallskip

\noindent McSween, H. 1976. A new type of chondritic meteorite found in lunar soil. {\it Earth and Planetary Science Letters} 31:2, 193-199.
\smallskip

\noindent Melosh, H. 1985. Ejection of rock fragments from planetary bodies.
{\it Geology}, 13, 144-148.
\smallskip

\noindent Melosh, H. 1988. The rocky road to panspermia. {\it Science}, 332, 687-688.
\smallskip

\noindent Melosh, H. 1989. {\it Impact Cratering: A Geologic Process}. Oxford University Press,
New York.
\smallskip

\noindent Melosh, H. J. 1993. Blasting rocks off planets. {\it Nature}, 363, 498-499.
\smallskip

\noindent Melosh, H. J., W. B. Tonks 1993.  Swapping rocks: ejection and exchange of surface material amoung the terrestrial planets. {\it Meteoritics}, 28, 398.
\smallskip

\noindent Mojzsis, S.J., G. Arrhenius, 
K. D. McKeegan, T. M. Harrison, A. P. Nutman, and C. R. L. Friend 1996. 
Evidence for life on Earth before 3,800 million years ago. {\it Nature},
340, 55-59.
\smallskip

\noindent Moorbath, S. 2001. A geochronological 
re-evaluation of claims for the contemporaneity of earliest life with major 
impacts. {\it Lunar and Planetary Science XXXII}, 1608.
\smallskip

\noindent Mustard, J. F. 1997. Constraints on the
magnitude of vertical and lateral mass transport on the Moon. 
{\it NASA Technical Report}, CR-1998-208203.
\smallskip

\noindent O'Keefe, J. and T. Ahrens 1986. Oblique impact: A process for obtaining
meteorite samples from other planets. {\it Science}, 234, 346-349.
\smallskip

\noindent Peterson, E., F. H\"orz, and S. Chang 1997.  Modification of amino acids
at shock pressures of 3.5 to 32 GPa. {\it Geochim. Cosmochim. Acta}, 61,
3937-3950.
\smallskip

\noindent Pierazzo, E., and C. Chyba 1999.
 Amino acid survival in large cometary impacts. {\it Meteoritics and 
Planetary Science},  34, 909-918.
\smallskip

\noindent Pierazzo, E. and H. Melosh 2000a.  Understanding oblique impacts from experiments, 
observations, and modeling. {\it Annu Rev Earth Planet Sci}, 28, 141-67.
\smallskip

\noindent Pierazzo, E. and H. Melosh 2000b. Hydrocode modeling of oblique impacts:
The fate of the projectile. {\it Meteoritics and Planetary Science}, 35, 117-130.
\smallskip

\noindent Rosing, M. 1999.
 13C-depleted carbon microparticles in $>3700$ Ma sea-floor sedimentary rocks 
 from West greenland. {\it Science}, 283, 674-676.
\smallskip

\noindent Schidlowski, M. 1988.
 A 3,800 million year isotopic record of life from carbon in sedimentary rocks.
 {\it Nature}, 333, 313-318.
\smallskip

\noindent Schopf, J. W. and B. M. Packer 1987. Early Archaean 3.3-billion to 
3.5-billion-year-old microfossils from Warrawoona group, Australia. 
{\it Science}, 237, 70-73.
\smallskip

\noindent Sleep, N., and K. Zahnle 2001.  Carbon dioxide cycling and implications for climate on ancient Earth. {\it JGR}, 106 E1, 1373-1400.
\smallskip

\noindent Sleep, N., K. Zahnle, J. Kasting, and H. Morowitz 1989. Annihilation of
ecosystems by large asteroid impacts on the early Earth. {\it Nature}, 342,
139-142.
\smallskip

\noindent Stadel, J. 2001. Cosmological N-body
Simulations and their Analysis. Ph.D. Thesis, Department of Astronomy,
University of Washington, Seattle.
\smallskip

\noindent Stoffler, D. 2000. Maskelynite confirmed as diaplectic glass: 
indication for peak shock pressures of 45 GPa in all Martian Meteorites.
{\it 31st Annual Lunar and Planetary Science Conference}, abstract no. 1170.
\smallskip

\noindent Stoffler, D., K. Keil, and E. Scott 1991. Shock metamorphism of
ordinary chondrites. {\it Geochim. Cosmochim. Acta}, 55, 3845-3867.
\smallskip

\noindent Thomas-Keprta, K., S. Clemett, D. Bazylinski, J. Kirschvink, D. McKay,
S. Wentworth, H. Vali, E. Gibson, M. McKay, and C. Romanek 2001.
Truncated hexa-octahedral magnetite crytsals in ALH84001: Presumptive
biosignatures. {\it Proceedings of the National Academy of Science USA}, 98(5)
2164-2169.
\smallskip

\noindent Tingle, T., J. Tyburczy, T. Ahrens, and C. Becker 1992. The fate of
organic matter during planetary accretion: Preliminary studies of the
organic chemistry of experimentally shocked Murchison meteorite. {\it Origins
of Life Evol. Biosph.}, 21, 385-397.
\smallskip

\noindent Tingle, T. 1998. Accretion and differentiation of carbon in the early
Earth. {\it Chemical Geology}, 147, 3-10.
\smallskip

\noindent Tyburczy, J., B. Frisch, and T. Ahrens, 1986. Shock-induced volatile loss
from a carbonaceous chondrite: Implications for planetary accretion. {\it Earth
and Planetary Science Letters}, 80, 201-207.
\smallskip

\noindent Valley, J. W., J. M. Eiler, C. M . Graham, E. K. Gibson, C. S. Romanek and E. M. Stolper, 1997. Low-temperature carbonate concretions in the Martian meteorite ALH84001: Evidence from stable isotopes and mineralogy. {\it Science},  275, 1633-1638.
\smallskip

\noindent Walker, J. C. G., and K. J. Zahnle 1986. Lunar nodal tide and distance to the Moon during the Precambrian. {\it Nature}, 320, 600.
\smallskip

\noindent Weiss, B. P., J. L. Kirschvink, F. J. Baudenbacher, H. Vali, N. T. Peters, F. A. Macdonald, and J. P. Wikswo 2000. A low temperature transfer of ALH84001 from Mars to Earth. {\it Science} 290, 791-795.
\smallskip

\noindent Williams, G. E. 2000. Geological constraints on the precambrian history of the Earth's rotation and the Moon's orbit  {\it Review of Geophysics}, 38(1), 37-59.
\smallskip

\noindent Zharkov, V.N. 2000. On the history of the lunar orbit, {\it Solar System 
Research}, 34(1), 1-11.
\smallskip

\newpage



\noindent {\bf Table 1.} The fraction of impactor mass delivered to the Moon for a range of Earth-Moon distances.  $\eta$ is the Earth-Moon distance in Earth radii and $v_{min}$  is the velocity of the shell's trailing edge.  The orbital period of the Moon and its interaction time with the shell are given, in days.  $F_{direct}$ is the fraction of the impactor mass that reaches the Moon.\\

\begin{tabular}{|c|c|c|c|c|}\hline

 $ \eta$ & $v_{min}, \ km \ s^{-1}$ & Orbital Period, days & Interaction Time, days & $F_{direct}$ \\ \hline \hline
21.6 & 10.94 & 5.9 & 0.30 & $2.0 \times 10^{-6}$  \\ \hline
26.5 & 10.99 & 8.0 & 0.40 & $1.2 \times 10^{-6}$  \\ \hline
31.3  & 11.02 & 10.2 & 0.50 & $7.9 \times 10^{-7}$  \\ \hline
36.2  & 11.05 & 12.7 & 0.58 & $5.3 \times 10^{-7}$  \\ \hline
41.0  & 11.06 & 15.3 & 0.66 & $3.9 \times 10^{-7}$  \\ \hline
45.9  & 11.08 & 18.1 & 0.78 & $2.9 \times 10^{-7}$  \\ \hline
50.7  & 11.09 & 21.1 & 0.84 & $2.2 \times 10^{-7}$  \\ \hline
55.5  & 11.10 & 24.1 & 0.91 & $1.7 \times 10^{-7}$  \\ \hline
60.3  & 11.11 & 27.3 & 0.99 & $1.4 \times 10^{-7}$  \\ \hline
\end{tabular}

\newpage


\noindent {\bf Table 2.} Impact results for orbital transfer.  $\frac{N_\oplus}{N_m}$ is the ratio of Terran to lunar impacts, from Eq. 8, $f_\oplus$ is the fraction of particles returning to the Earth, and $f_m$ is the fraction of particles returning to the Moon.  Also listed are the expected number of lunar impacts from the scaling arguments, the lunar transfer efficiency, $F_e$, and the fraction of the impactor mass transferred to the Moon via this method, $F_{orb}$. \\

\begin{tabular}{|c|c|c|c|c|c|c|c|c|}\hline
$v_{\infty}$ & Earth & Lunar & $\frac{N_\oplus}{N_m}$ & 
$f_\oplus$ & $f_m$  & Scaled  & $F_e$ & $F_{orb}$ \\
$km \ s^{-1}$ & impacts & impacts &  & 
 &  & lunar impacts &  &  \\  \hline \hline
0.0 & 138 & 1 &140.0 & 0.61  & 0.004  & 0.98 & $4\times10^{-3}$ &  $3\times10^{-7}$\\  \hline
1.0 & 14 & 0 &130.4 & 0.06 &  0.0 & 0.11 &  $5\times10^{-4}$ & $5\times10^{-7}$ \\  \hline
1.8 & 5 & 1 & 113.3 & 0.02 &  0.004 &  0.04 &   $2\times10^{-4}$  & $2\times10^{-7}$ \\  \hline
2.3 & 1 & 0 & 101.5 & 0.004 & 0.0  &  0.01 &  $4\times10^{-5}$ & $2\times10^{-8}$ \\  \hline
2.7 & 1 & 0 & 92.4 &  0.004 & 0.0  &  0.01 &  $4\times10^{-5}$ & $4\times10^{-6}$ \\  \hline
3.3 & 1 & 0 & 80.0 &  0.004 & 0.0  &  0.01 &   $4\times10^{-5}$ & $5\times10^{-6}$\\ \hline

\end{tabular}

\newpage


\noindent {\bf Table 3.}  Lunar impact velocities.  Listed are the most likely impact velocities for an angle less than or equal to $\theta$, the probability that a given impact is less than or equal to $\theta$, and the corresponding impact pressures experienced by the impactor on the Moon. \\

\begin{tabular}{|c|c|c|c|}\hline
$v_{exp} \ km \ s^{-1}$ & $\theta$, degrees & $P\left(\le\theta\right)$ & Impact Pressure, GPa \\ \hline \hline
5.0 & 90 & 1.00 &  48 \\ \hline
2.5 & 77 & 0.95 & 12 \\ \hline
2.0 & 64 & 0.81 & 8 \\ \hline
1.0 & 45 & 0.50 & 2 \\ \hline
0.5 & 35 & 0.33 & 0.5 \\ \hline
0.1 & 20 & 0.12 & 0.02 \\ \hline
 
\end{tabular}

\newpage


\noindent {\bf Table 4.} Values of the fit parameters for the lunar cratering record. \\

\begin{tabular}{|c|c|c|c|c|c|c|}\hline
$\alpha$ & $\beta$   & $\gamma$ & $\tau$, Gyrs  & $\mu$, Gyrs & $\tau_c$, Gyrs & $\chi^2$ \\ \hline \hline 
$2.24 \times 10^{-5}$ & $9.82 \times 10^{-11}$ & - & 0.144 & - & - & 0.50  \\ \hline
$2.26 \times 10^{-5}$ & $3.57 \times 10^{-12}$ & $1.27 \times 10^{-3}$ &
	 0.129 & 3.9 & 0.07 & 0.99  \\ \hline

\end{tabular}
\newpage


\noindent {\bf Table 5.} Ejected material as fractions of the impactor mass calculated for the different velocity and ejecta pressure regimes.  $v_i$ is the impactor velocity, $M_e$ is the ejecta mass, and $M_i$ is the impactor mass. \\

\begin{tabular}{|c|c|c|c|c|}\hline
$v_i$ & 
$M_e/M_i$ &
$M_e/M_i$  & 
$M_e/M_i$ & 
$M_e/M_i$ \\ 
$\left( km \ s^{-1}\right)$ &
1 GPa &
10 GPa &
30 GPa &
70 GPa \\ \hline \hline  
14  & no escape & no escape & no escape & no escape \\ \hline
22.5  & $1.45\times10^{-5}$ & $1.45\times10^{-4}$ & $4.34\times10^{-4}$ & $1.01\times10^{-3}$  \\ \hline
25 & $3.51\times10^{-4}$ & $3.51\times10^{-3}$ & $1.05\times10^{-2}$ & $2.46\times10^{-2}$  \\ \hline
30 & $9.14\times10^{-4}$ & $9.14\times10^{-3}$ & $2.74\times10^{-2}$ & $6.40\times10^{-2}$ \\ \hline
40  & $1.78\times10^{-3}$ & $1.78\times10^{-2}$ & $5.34\times10^{-2}$ & $1.25\times10^{-1}$  \\ \hline
50 & $2.46\times10^{-3}$ & $2.46\times10^{-2}$ & $7.38\times10^{-2}$ & $1.72\times10^{-1}$ \\ \hline
65  & $3.30\times10^{-3}$ & $3.30\times10^{-2}$ & $9.89\times10^{-2}$ & $2.31\times10^{-1}$ \\ \hline

\end{tabular}
\newpage


\noindent {\bf Table 6.} Of the Terran material on the Moon, the mass fraction from a given impact that arrived there under conditions conducive to the survival of molecular fossils, organics, significant volatile inventories, or isotopes.  The fractions of material with lunar impact velocities of $\le 5 \ km \ s^{-1}$, $\le 2 \ km \ s^{-1}$, and $\le 0.5 \ km \ s^{-1}$ are 1.0, 0.81, and 0.33 respectively (see Table 3). \\

\begin{tabular}{|c|c|c|c|c|}\hline
$v_i$ &
Biomarkers   & 
Organics &
Volatiles  & 
Isotopes \\
$km \ s^{-1}$ &
\% &
\% &
\% &
\% \\ \hline \hline
14  & no escape & no escape & no escape & no escape \\\hline
22.5 & 0.002& 0.05 & 0.14 & 0.40 \\ \hline
25  & 0.04 & 0.99 & 2.96 & 8.53 \\ \hline
30 & 0.09 & 2.06 & 6.19 & 17.8 \\ \hline
40 & 0.12 & 2.84 & 8.53 & 24.6 \\ \hline
50  & 0.12 & 3.01 & 9.02 & 26.0 \\ \hline
65  & 0.12 & 2.94 & 8.83 & 25.4 \\ \hline

\end{tabular}
\newpage

\newpage
\noindent{\Large \bf FIGURE CAPTIONS:}
\bigskip\bigskip

\noindent{\large Figure 1.} \hspace{4mm}
 The cumulative number of impacts as a function of time for the 
simulation with $v_\infty = 0.0 \ km \ s^{-1}$.  Most of the impacts
(123 out of 138) occur within the first 100 years, and the rate  
tapers off substantially after 1000 years.
\bigskip

\noindent{\large Figure 2.} \hspace{4mm}
 Analytical fits to the BVSP dataset for two models of the
time evolution of impactor flux.  The dashed line is the lunar cataclysm, 
and the solid line is the exponential model.  The plot
is the cumulative crater density per square kilometer 
(in both size and time) for craters
larger than 4 km shown with the BVSP data.  The
effects of the lunar cataclysm can be seen around 3.9 Ga.
\bigskip

\noindent{\large Figure 3.} \hspace{4mm}
 Plot of the total regolith depth accreted as a 
function of time.  The lines correspond to the models as stated in
Fig. 2
\bigskip

\noindent{\large Figure 4.} \hspace{4mm}
 Plot of the fractional volume of Terran material as a function
of time in the lunar regolith. The lines correspond to the models as stated in
Fig. 2
\bigskip
\newpage






\end{document}